\documentclass{emulateapj}

\def\be{\begin{equation}}
\def\ee{\end{equation}}

\def\arcdeg{{^{\circ}}}%

\def\bh{M_{\rm BH}}
\def\bhsp{M_{\rm BH,1}}
\def\bhs{M_{\rm BH,2}}
\def\mbh{m_{\rm BH}}

\def\pc{\rm ~pc}
\def\mpc{\rm ~mpc}
\def\msun{M_{\odot}}
\def\rtids{r^{\rm s}_{\rm tid}}
\def\rtidb{r^{\rm b}_{\rm tid}}
\def\rsun{R_{\odot}}
\def\ab{a_{\rm b}}
\def\aeff{a_{\rm eff}}
\def\ah{a_{\rm h}}
\def\abbh{a_{\rm BBH}}
\def\AU{\rm ~AU}
\def\kms{\rm ~km~s^{-1}}
\def\kpc{\rm ~kpc}
\def\yr{\rm ~yr}

\shorttitle{Hypervelocity binary stars}
\shortauthors{Lu, Yu, \& Lin}

\begin{document}
\title{Hypervelocity binary stars: smoking gun of massive binary 
black holes 
}
\author{Youjun Lu\altaffilmark{1}, Qingjuan Yu\altaffilmark{1}, \& D.\ N.\ C.\ Lin\altaffilmark{1,2}
} 
\altaffiltext{1}{Department of Astronomy and Astrophysics, University of California, Santa Cruz, CA 95064, USA}
\altaffiltext{2}{Kavli Institute of Astronomy and Astrophysics, Peking University, Beijing, China
}
\email{lyj, yqj, lin@ucolick.org}

\begin{abstract}

The hypervelocity stars recently found in the Galactic halo are
expelled from the Galactic center through interactions between binary
stars and the central massive black hole or between single stars and a
hypothetical massive binary black hole. In this paper, we demonstrate
that binary stars can be ejected out of the Galactic center with
velocities up to $10^3\kms$, while preserving their integrity, through
interactions with a massive binary black hole. Binary stars are
unlikely to attain such high velocities via scattering by a single
massive black hole or through any other mechanisms.  Based on the
above theoretical prediction, we propose a search for binary systems
among the hypervelocity stars. Discovery of hypervelocity binary
stars, even one, is a definitive evidence of the existence of a
massive binary black hole in the Galactic center.

\end{abstract}

\maketitle

\section{Introduction}\label{sec:intro}

Evidence of a massive black hole (MBH) is securely established in the
Galactic center (GC) \citep{Schodel02,Ghez03,Ghez05,Eisenhauer}.
Recent discoveries of young stars in the sub-parsec region around the
GC \citep{Sanders,Ghez03} and hypervelocity stars (HVSs; first
recognized by \citealt{Hills88}) in the Galactic halo
\citep{Brown05,Brown06a,Brown06b,Edelmann05,Hir05} have stimulated the
hypothesis of a second MBH or an intermediate-mass BH (IMBH) which may
be common in galactic nuclei if they are assembled through mergers of
smaller galaxies and/or star clusters with central MBHs or IMBHs
\citep{BBR80,Y02,VHM03}. IMBHs can transport young stars to their
current locations \citep{HM03} where {\it in-situ} star formation is
suppressed by the strong tidal force of the MBH.  Through close
encounters, a massive binary black hole (BBH) may also eject nearby
stars to become HVSs \citep{YT03,BGP05,Levin06,SHM06}. Direct
observational evidences for a secondary MBH are difficult to establish
because the semimajor axis of the BBH could be well within the orbit
of the most central `S' stars which have longer-than-a-decade orbital
periods \citep{Schodel02,Ghez03}.

With a simple scaling argument in \S~\ref{charadii} and a series of
numerical calculations in \S~\ref{method} and \S~\ref{results}, we
demonstrate that hypervelocity binary stars (HVBSs) with velocities up
to $10^3 \kms$ can be ejected out only by their dynamical interactions
with a massive BBH.  We propose a search for HVBSs among
HVSs. Discovery of HVBSs, even one, in the Galactic halo will be a
definitive evidence for the existence of a massive BBH in
the GC.

\section{Domains of relevant dynamical processes}
\label{charadii}

\begin{figure} \epsscale{0.8}
\plotone{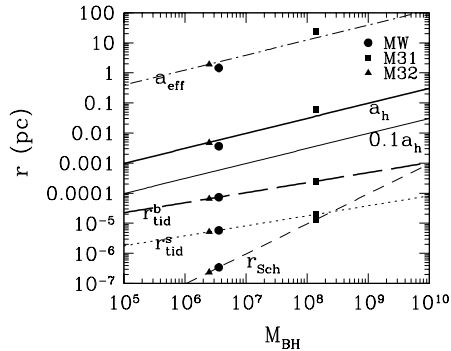}
\caption{Various characteristic radii in galactic centers with massive
black holes: (1) the Schwarzschild radius (dashed line); (2) the tidal
radius for a single star with solar mass and solar radius (dotted
line); (3) the tidal breakup radius for a binary star with semimajor
axis $\ab=0.1\AU$ (long-dashed line; $\eta=2.21$ is adopted here, see
text); (4) the radius of the sphere of influence of central MBH
(dot-dashed line); and (5) the transition radius for a hard binary
black hole with mass ratio of $\nu=0.01$ (thick solid line), the thin solid line represents
$\abbh=0.1\ah$.  The characteristic radii for the MBHs in the Milky
Way (with $\bh=3.6\times 10^6\msun$ and $\sigma=103\kms$),
M31 ($1.4\times 10^8 \msun$, $160\kms$), and M32
($2.5\times 10^6\msun$, $ 75\kms$) are labeled as
circles, squares and triangles, respectively. } \label{fig:f1}
\end{figure}

We characterize the dynamics of stars around a MBH in galactic centers
with the following critical radii: (1) the Schwarzschild radius is
$r_{\rm Sch}=2G\bh/c^2=3.4\times 10^{-7}\mbh~\pc$, where $\mbh$ is the
MBH's mass $\bh$ normalized by that of the MBH in the GC (in unit of
$3.6 \times 10^6\msun$; \citealt{Ghez05,Eisenhauer}), $G$ is the
gravitational constant and $c$ is the speed of light; (2) the tidal
radius for a single star with mass $m_*$ and radius $r_*$ is
$\rtids=r_*\left(\eta^2{\bh}/{m_*}\right)^{1/3}= 3.5\times 10^{-6}
(\eta^2\mbh)^{1/3} ({\msun}/{m_*})^{1/3} ({r_*}/{\rsun})\pc$, where a
constant $\eta =2.21$ and 0.844 for a homogeneous, incompressible body
and an $n=3$ polytrope respectively \citep{ST92,Diener95}; (3) the
tidal breakup radius for a binary star (with semimajor axis $\ab$ and
component masses $m_1$ and $m_2$, respectively) is $\rtidb=
\ab[2\bh/(m_1+m_2)]^{1/3}= 7.4\times 10^{-5}
\mbh^{1/3} (\ab/0.1\AU)[2\msun/(m_1+m_2)]^{1/3}\pc$;
(4) the radius of the sphere of influence of the MBH is
$\aeff={G\bh}/{\sigma^2}\simeq 2.3~\mbh^{0.5}\pc$, where $\sigma$ is the
1D velocity dispersion of the spheroidal component of the galaxy and
the $\bh-\sigma$ relation ($\bh\propto\sigma^{4.02}$;
\citealt{Tremaine02}) is used to replace $\sigma$ by $\bh$; and (5) a
massive BBH becomes ``hard'' when its semimajor axis $\abbh\la \ah$,
where the critical value $\ah={G\bhs}/{4\sigma^2}\simeq 5.9\times
10^{-3}~\mbh^{0.5} ({\nu}/{0.01}) \pc $, $\nu=\bhs/\bh$,
$\bh=\bhsp+\bhs$ ($\bhsp\ge\bhs$), and $\bhsp$ and $\bhs$ are the
masses of the primary and secondary MBHs, respectively
\citep{Quinlan96}.  BBHs, if exist in nearby normal galactic centers,
would spend most fraction of their lifespan at $\abbh \sim
0.01\ah-1\ah$ if $10^6\msun\la\bh\la 10^8\msun$ \citep{Y02}, so that
for companions around these MBHs with $\nu\ga 10^{-2}$, we generally
have $r_{\rm Sch}<\rtids <\rtidb\la\abbh\la\ah<\aeff$ (see
Fig.~\ref{fig:f1}).

With these definitions, the vicinity of a MBH can be partitioned by
various dominant physical processes.  Most of the stars within $r<\aeff$
have negative energy and are bound to the central MBH and their
kinematic properties have been used to determine the MBH mass
\citep{Richstone}. Stars that enter into the region $r\la r_{\rm Sch}$
(if $\rtids<r_{\rm sch}$) are swallowed as a whole by the MBH.  Stars
that enter into the region with $r_{\rm Sch}<r\la\rtids$ are tidally
disrupted. A fraction of the disrupted star is accreted by the MBH
through a disk-like flow with outbursts of enormous radiation and the
other fraction may be ejected out of galactic nuclei
\citep{Hills75,Rees88}.

Here, we are primarily interested in the origin of HVSs.  A star
initially unbound or weakly bound to a MBH will attain a velocity
$v\simeq\sqrt{G\bh/r}\simeq 5.5\times10^3\kms\mbh^{1/2}
(1\mpc/r)^{1/2}$ upon moving within a distance $r$ from the MBH, where
$\mpc\equiv 10^{-3}\pc$. If the star then ensures a velocity
perturbation $\delta v\ll v$ with an increase in the specific energy
of the star $v\delta v$ which is much larger than the magnitude of its
initial specific energy ($|v^2/2-{G\bh}/{r} |$), it would escape from
the MBH with velocity $\sim \sqrt{2v\delta v}$. For example, a binary
star that enters into the region $\rtids<r\la\rtidb$ is likely to be
tidally broken up, during which episode, each star would receive a
$\delta v$ comparable to its binary orbital velocity.  Consequently,
one component of the binary star would be ejected out of the system
with hypervelocity up to $10^3\kms$ and the other would become more
tightly bound to the BH \citep{Hills88}.  Around a hard BBH with
$\abbh\la\ah$, most low-angular-momentum single stars that enter into
the region $\rtids<r\la\abbh$ can also be ejected by the BBH.
The r.m.s. of the velocities of the ejected stars at infinity is given
by
\begin{eqnarray}
v^{\infty}_{ej} & \simeq & \sqrt{2KG\bhsp\bhs/(\bh\abbh)} \nonumber \\
                    & \sim   & 900 \kms \mbh^{0.25} (1-\nu)^{1/2}
                               (0.1\ah/\abbh)^{1/2},
\label{eq:1}
\end{eqnarray}
(see eq.~17 in \citealt{Y02}), where $K\simeq1.6$ is a constant and
the $\bh-\sigma$ relation obtained by Tremaine {\it et al.} (2002) is
adopted.
The main portion of $\delta v$ is due to the impulse induced by the
secondary MBH $F\delta t$, where $F\sim G\bhs/\abbh^2$ is the force
per unit mass from the secondary BH and $\delta t\sim \sqrt{G\bh/\abbh^3}$ 
is the interaction time. Provided that $\abbh$
is substantially larger than $\rtidb$, there is a large
probability for a binary star's periapse not to enter into the region
$r\la \rtidb$ (with respect to either component of the BBHs) before it
is ejected.  Consequently, binary stars which enter into the region
$\rtidb<r\la \abbh$ may be ejected, with large velocities while
preserving their systems' integrity, as HVBSs.

\section{Numerical method}
\label{method}
In order to determine the probability of HVBS formation, we simulate
the interaction of binary stars with BBHs.  Instead of carrying out a
comprehensive series of prohibitively complex four-body problem, we
note that the motion of a binary star with masses $m_1$ and $m_2$ in
the potential of a BBH may be greatly simplified in the limit of
$m_1,m_2\ll\bhsp,\bhs$ because the motion of the BBH is not affected
by the binary star and can be described in terms of a two-body
problem. We set the center of mass of the BBH is set to be at rest at
the origin of the coordinate system. Each component of the binary star
moves in the potential of the rotating BBH and that of its stellar
companion ({\it i.e.}, $Gm_j/|\vec{r}_j-\vec{r}_i|- Gm_j
\vec{r}_i\cdot\vec{r}_j/r_j^3$, $i\ne j$, $i,j=1,2$, where
$\vec{r}_i$ and $\vec{r}_j$ are the position vectors of the two
components).  Here Newtonian mechanics is applied.  The initial
velocity of the center of mass of the binary star at infinity is set
to be the velocity dispersion $\sigma$ of the galaxy.  We choose its
impact direction randomly. The impact parameter $b$ is chosen so that
the pericenter distance $r_p$ of its corresponding Keplerian orbit
[which is related with $b$ through $b=r_p(1+{2G\bh}/{(\sigma^2
r_p}))^{1/2}$] is in the range from $0.1\abbh$ (or $0.01\abbh$) to
$2\abbh$.  The angular momentum vector and the initial phase
of the binary star's orbits about their center of mass are chosen
randomly. For each set of initial conditions, we integrate a system of
differential equations with an explicit 5(4) order Runge-Kutta
scheme. The center of mass of the binary star is analytically
extrapolated from $r=\infty$ to $100\abbh$ along a Keplerian orbit
about a point mass $\bh$ where the numerical integration starts.

Some integrations are time consuming because the binary is captured into
weakly-bound orbits to the BBH and makes many revolutions before it is
expelled. We adopt similar approximations to expedite the numerical
integrations as that introduced by \citet{Quinlan96} for the studies of
interactions between a single star and a BBH.

The fate of the interaction of a binary star with a massive BBH mainly
falls into two cases: (1) the binary star is ejected as a whole from
the BBH with some changes in its intrinsic orbital semimajor axes and
eccentricities; or (2) the binary star is dissociated by its interactions
with the BBH and both dissociated components are ejected with
hypervelocities.

\section{Numerical results and discussions}
\label{results}

\begin{figure} \epsscale{1.1}
\plotone{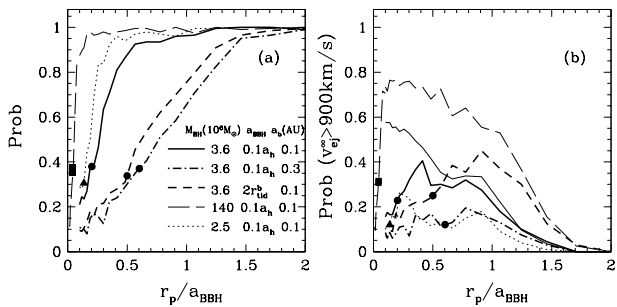}
\caption{Probability for ejecting hypervelocity binary stars.
(a) Fraction of undisrupted binary stars that are expelled
after their interactions with a hypothetical BBH.  The BBH
mass ratio is set to $\nu=0.01$. The thick lines are for the BH mass
in the Milky Way with different model parameters for the BBH and
binary star's semimajor axes.  The thin long-dashed line is for
the BH mass in M31 and the thin dotted line is for M32. The
circles/triangles/squares mark the radii $\rtidb$ for each cases. 
(b) Fraction of binary stars that are expelled away
as a whole with hypervelocity $v^{\infty}_{ej}>900\kms$ after their
interactions with the assumed BBH in galactic centers.  The line types
and the solid points have the same meanings as those in (a). As a
reference, the thin solid line represents the fraction of single
stars that are expelled with $v>900\kms$ after their three-body
interactions with an identical BBH as that represented by the thick
solid line. 
} \label{fig:prob}
\end{figure}

\begin{figure} \epsscale{0.9}
\plotone{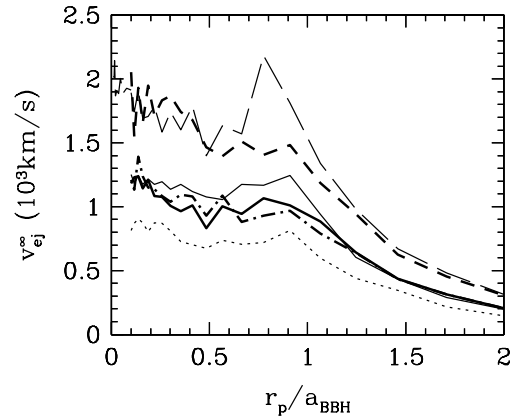}
\caption{The r.m.s.\ velocity of the ejected binary stars at infinity versus
the pericenter distances of their initial injection orbits.
The line types have the same meaning as those in Fig.~\ref{fig:prob}.}
\label{fig:vej}
\end{figure}

We present numerical results of the interaction of a binary star with
a massive BBH in Figures \ref{fig:prob} and \ref{fig:vej}.  We assume
various values for the semimajor axes of binary stars (with $\ab=
0.1\AU$ and $0.3\AU$; note that here $0.3\AU$ is taken as an upper
limit because binaries with larger $a_b$ are likely to be disrupted by
encounters with single stars in the GC within a Hubble time,
\citealt{YT03}) and for the BBH semimajor axes: $\abbh=0.1\ah
=0.37,5.88,0.48\mpc $ [$= (4.9, 23, 7.2)~\rtidb$ for the Milky Way,
M31, and M32, respectively] and $\abbh=2\rtidb =0.15\mpc$ (for the
Milky Way).  The component masses of the binary star are set to be
$m_1=m_2=1\msun$ and our results are not modified by other values of
binary masses. The BBH mass ratio is set to be $\nu=0.01$ unless
otherwise stated. For simplicity, the BBH is set to be on a circular
orbit and the initial intrinsic orbit of binary stars are also
circular. Figure \ref{fig:prob}(a) shows that the probability of the
binary star being ejected away as a whole from the BBH is generally
larger than 50\% if $r_p>\rtidb$ and that binary stars are more likely
to be dissociated with decreasing pericenter distances $r_p$. Given
the same BBH mass and semimajor axis, binary stars with larger
$\ab$'s are more likely to be dissociated (see the thick solid line
and dot-dashed line).  If we keep $\abbh =0.1\ah$ but set $\nu=0.001$
for the assumed BBH in the GC, the probability of ejecting HVBSs is
smaller (down to 20\% at $r_p\simeq\abbh$) since $\abbh$ is smaller
than the tidal radius of the primary MBH $\rtidb$ and most of the
binary stars are tidally broken up. 

Figure~\ref{fig:vej} shows the dependence of the ejection velocities
of HVBSs on the pericenter distance $r_p$, which is generally
consistent with the dependence of the ejection velocities of single
stars through three-body interactions with the BBH (thin solid line).
The $v^{\infty}_{ej}$ can attain values up to $10^3\kms$ at
$r_p\sim\abbh=0.1\ah$ and it remains roughly constant with
decreasing $r_p$ but decreases with increasing $r_p$. The ejection
velocities of HVBSs depend on the BBH mass and semimajor axis (see
eq.~\ref{eq:1}). With all else being equal, the ejection velocities
increases with increasing BH mass (see the thin lines for M31 and M32,
respectively, in Fig.~\ref{fig:vej}); and decrease with increasing BBH
semimajor axes (see thick short-dashed line and thick solid line).

Figure \ref{fig:prob}(b) shows the probability of the binary star
being ejected away as a whole with hypervelocity $v^{\infty}_{ej}
>900\kms$ from the BBH has a peak value of 10\%--50\% at $r_p\sim
(0.1$-$1.5)\abbh$ and it decreases for both larger and smaller values
of $r_p$.  (Due to the deceleration in the Galactic potential, an
ejected binary star, HVBS, with
$v^{\infty}_{ej}=900\kms$ would attain an asymptotic velocity
$700\kms$ at the Galactocentric distance $50\kpc$, which is
corresponding to the observed radial velocities of HVSs in the
Galactic halo.)
The decline of ejection probability of binary stars with high
velocities at large $r_p$ is consistent with that for single stars
(see thin solid line). But the rarity of ejecting HVBSs at small $r_p$
is mainly due to the dissociation of binary stars by the Hills
mechanism as indicated in Figure \ref{fig:prob}(a).  (Note that the
encounters with a BBH at small $r_p$ require low-angular-momentum
orbits.  After the depletion of such an initial population, the
likelihood to eject HVBSs at small $r_p$ declines because, through
two-body relaxation, the remaining binary stars would slowly lose
their angular momenta to other stars and reach the vicinity of the BBH
preferentially with large $r_p$.) Our calculations show that for a
larger BBH semimajor axis $\abbh=0.3\ah$ in the GC, the peak
probability is about 5\%--10\%, which is still non-negligible.  The
ratio of number densities between the HVBSs and the single HVSs
contains information on the fraction of binary stars in the GC, the
BBH mass ratio and semimajor axis.  If a BBH with $\nu\sim 0.01$ and
$\abbh\sim 0.1\ah$ exists in the GC, according to Figure
\ref{fig:prob}(b), the probability of binary stars being ejected with
$v^{\infty}_{ej}>900\kms$ would be at least about $1/2-1/3$ of the
probability of single stars being ejected with such hypervelocities
provided $r_p\ga \rtidb$. If the density of binary stars with $\ab
\sim (0.1-0.3)\AU$ is about 10\% of that of the single stars in the
GC, the fraction of HVBSs among the HVSs would then be at least about
$3\%-5\%$.

The ejection rate of HVSs by an assumed BBH (with $\nu=0.01$ and
$\abbh=0.5\mpc$)
\footnote{The hardening timescale of the BBH due to three-body
interactions with stars passing through its vicinity can be $\ga
10^9\yr$, and the gravitational radiation is the dominant mechanism
to make the BBH lose energy with a timescale of $10^8\yr$. According
to Figure 2 in \citet{YT03}, the BBH should have $\nu\ga10^{-3}$
in order to eject stars with $v_{ej}^\infty\ga700\kms$ (which corresponds
to a velocity $\ga400\kms$ at the Galactocentric distance $50\kpc$) 
and to have the present accretion rate of BH companions is not
unusually large compared to the average over the lifetime of the Galaxy.
} in the GC
would be $\sim 10^{-4}\yr^{-1}$ if the low-angular-momentum stars
initially in the loss cone are depleted and the loss cone is refilled
by two-body relaxation processes \citep{YT03}.  If the fraction of
binary stars with $\ab\la 0.3\AU$ in the GC is about 10\% of total
stars, the ejection rate of HVBSs would be $\ge 5\times
10^{-6}\yr^{-1}$. By applying the same method that is adopted by
\citet{YT03}, we obtain ejection rates for M32 (with $\nu=0.01$ and
$\abbh=0.48\mpc$), which is similar to that for the Milky Way.  But
for M31 (with $\nu=0.01$ and $\abbh=5.9\mpc$), we obtain lower
ejection rates ($\sim 10^{-5}\yr^{-1}$ for HVSs and $\sim
10^{-6}$--$10^{-7} \yr^{-1}$ for HVBSs).  For M31, we adopt the
surface density distribution estimated by \citet[][eq.~1]{KB99} and we
neglect the departure of stellar distribution from spherical symmetry
\citep{Bender05} as for the other cases.  Asphericity, if significant,
would enhance the ejection rates of HVSs and HVBSs.

Several HVSs have recently been discovered through spectroscopic
observations \citep{Brown05,Brown06a,Brown06b,Edelmann05,Hir05}.  The
presence of any HVBSs among this sample is difficult to be spatially
resolved. But the spectrum of the more conspicuous primary would
exhibit measurable ($\sim 100 \kms\sin i$ ) periodic variations on the
timescale of a few to several ten days.  Transits may also introduce
observable periodic light curve modulations.  Non-detection of HVBSs
does not rule out the existence of a BBH in the GC since their absence
can be attributed to: (1) a low rate of binary-stars injection into
the region $\rtidb <r \la \abbh$; (2) a relative compact BBH with
$\abbh<\rtidb$; or (3) a relatively soft BBH which is unable to eject
stars with sufficiently high velocities.

In contrast, the detection of any HVBSs unequivocally proves the
existence of BBH because the production of HVBSs by any other
mechanism appears to be challenging and unlikely. (1) The Hills mechanism
cannot
lead to the production of HVBSs because it requires the breakup of a
binary star by a single MBH. (2) In principle, tidal disruption of a
hierarchical triple star (a close binary plus a less tightly bound
tertiary companion) by the central MBH may lead to the capture of the
tertiary by the MBH and the ejection of the close binary as a
HVBS. However, the rate of ejecting HVBSs by this mechanism is
negligible if the following factors are taken into account. (a) The
fraction of the binaries with additional companions (triple,
quadruple, or quintuple) is about 60\% of the binaries observed
locally by \citet{Tokovinin06}. (b) All known tertiary companions
around the binaries have periods longer than $2\yr$ (with semimajor
axis $\ga 2\AU$) and $\ga$90\% of the tertiary companion have orbital
periods ranging from $5\yr$ to $10^5\yr$.
(c) The typical ejection
speed of the close binary due to the breakup of a hierarchical
triple with semimajor axis $2\AU$ is only $\sim400\kms$. If the
ejection speed follows a Gaussian distribution with a dispersion which
is about 20\% of the average speed (the dispersion is mainly due to
the different orbital phases and orientations of the binary orbit
relative to the MBH; e.g., see Fig.\ 1 in \citealt{Bromley06}), stars
with velocity $v_{ej}^\infty$ higher than $700\kms$ or $900\kms$ would
be beyond $3\sigma$ or $5\sigma$ from the mean, which corresponds to
small probabilities of $1.3\times10^{-3}$ or $3\times10^{-7}$.  Thus,
the ejection rate of HVBSs due to the tidal breakup of hierarchical
triple stars is only a factor of $0.6\times0.1\times1.3 \times 10^{-3}
\sim8\times10^{-5}$ or $2\times10^{-8}$ of the number of HVSs due to
Hills mechanism. We note that the argument above has some
uncertainties because the estimate derived for the solar neighborhood
is uncertain and it is not clear whether the triple distribution at
the Galactic center is the same as that in the solar neighborhood. (3)
HVSs may also be ejected as stars are scattered off a cluster of
stellar-mass black holes (SBHs) orbiting a central MBH.  In principle,
the ejected stars can be binaries.  But, the perturbation on the
velocity of ejection star(s), $\delta v$ is primarily due to close
encounters with slightly-more-massive SBHs at a distance about the
solar radius \citep{OL06}. Such close encounters can lead to
either the disruption of the binary system or an exchange in which the
SBH captures one component of the initial binary while the other is
ejected as a single star.  The newly formed star-SBH binary cannot be
ejected out of the GC as a HVBS since the SBHs are closely bound to
the central MBH with a substantially negative specific energy ($|E|\gg
v\delta v$ because usually $\delta v\ll v$).  Whereas the detection of
a HVS provides support for the existence of a MBH in the GC
\citep{Hills88,Brown05}, the detection of a HVBS will be the
smoking gun for the existence a massive BBH in the GC (or M31, M32).

Our numerical calculations show that both $\ab$ and the eccentricities
of the binary may be strongly modified during their interactions
with the BBH. The eccentricity of the binary-stars' orbits can be
excited up to 0.9--1, which may lead to the merger of the two
components of the binary star with small relative velocity. Therefore,
the interactions between binary stars on bound orbits and a BBH may
provide a channel for the formation of the S-stars discovered in the
Galactic center. It is plausible that both the formation of S-stars
and the ejection of HVSs are the by-products during the process of a
MBH inspiralling inward towards the primary MBH.

The velocity of some binary stars ejected by interactions with a BBH can be
smaller than the escape velocity of the Galactic halo and these stars are bound
to the Galaxy.  Such a binary may already be contained in existing
observational data. For example, Scorpius X-1, a X-ray binary, is found to have
a velocity of $\sim 480\kms$ with its past perigalactic distance of $\sim
500\pc$ to the GC \citep{MR03}.  This object could be explained as a binary
formed in the GC and ejected out by a BBH from there, though other
explanations, such as a natal kick of supernovae explosion, are possible.

\acknowledgments
This work is supported by NASA (NAG5-12151, NNG06-GH45G), JPL (1270927), 
NSF(AST-0507424).


\begin{thebibliography}{}

\bibitem[Baumgardt et al.(2005)]{BGP05}
Baumgardt, H., Gualandris, A., \& Portegies Zwart, S.\ 2005,
\mnras, 372, 174

\bibitem[Begelman et al.(1980)]{BBR80} Begelman,
M., Blandford, R.\ D., \& Rees, M.\ J.\ 1980,
\nat, 287, 307

\bibitem[Bender et al.(2005)]{Bender05} Bender, R., et al.\ 2005, 
\apj, 631, 280

\bibitem[Bromley et al.(2006)]{Bromley06} Bromley, B.\ C., Kenyon, S.\ J., Geller, M.\ J., Barcikowski, E., Brown, W.\ R., Kurtz, M.\ J.\ 2006, ApJ, 653, 1194	

\bibitem[Brown et al.(2005)]{Brown05}Brown, W.\ R., Geller, M.\ J.,
Kenyon, S.\ J., \& Kurtz, M.\ J.\ 2005, \apjl, 622, L33

\bibitem[Brown et al.(2006a)]{Brown06a}Brown, W.\ R., Geller, M.\ J.,
Kenyon, S.\ J., \& Kurtz, M.\ J.\ 2006a, \apjl, 640, L35

\bibitem[Brown et al.(2006b)]{Brown06b} Brown, W.\ R., Geller, M.\ J.,
Kenyon, S.\ J., \& Kurtz, M.\ J.\ 2006b, ApJ, 647, 303

\bibitem[Diener et al.(1995)]{Diener95}Diener, P., Kosovichev, A.\ G.,
Kotok, E.\ V., Novikov, I.\ D., \& Pethick, C.\ J.\ 1995,
\mnras, 275, 498

\bibitem[Edelmann et al.(2005)]{Edelmann05} Edelmann, H., Napiwotzki,
R., Heber, U., Christlieb, N., \& Reimers, D.\  2005,
\apjl, 634, L181

\bibitem[Eisenhauer et al.(2005)]{Eisenhauer} Eisenhauer, F., et al.\ 2005,
\apj, 628, 24

\bibitem[Ghez et al.(2003)]{Ghez03} Ghez, A.\ M., et al.\ 2003,
\apjl, 586, L127

\bibitem[Ghez et al.(2005)]{Ghez05} Ghez, A.\ M., et al.\  2005,
\apj, 620, 744

\bibitem[Hansen \& Milosavljevic(2003)]{HM03}Hansen, B.\ M.\ S., \&
Milosavljevic, M.\ 2003, \apjl, 593, L77

\bibitem[Hills(1975)]{Hills75}Hills, J.\ G.\ 1975, \nat, 254, 295

\bibitem[Hills(1988)]{Hills88}Hills, J.\ G.\ 1988, \nat, 331, 687

\bibitem[Hirsch et al.(2005)]{Hir05}Hirsch, H.\ A., Heber, U., 
O'Toole, S.\ J., \& Bresolin, F.\ 2005, \aa, 444, L61

\bibitem[Kormendy \& Bender(1999)]{KB99} Kormendy, J., \& Bender, R.\
1999, \apj, 522, 772

\bibitem[Levin(2006)]{Levin06} Levin, Y.\ 2006, \apj, 653, 1203

\bibitem[Mirabel \& Rodrigues(2003)]{MR03} Mirabel, I.\ F., \&
Rodrigues, I.\ 2003, A\&A, 398, L25

\bibitem[O'Leary \& Loeb(2006)]{OL06} O'Leary, R.\ M., \&
Loeb, A.\ 2006, astro-ph/0609046

\bibitem[Quinlan(1996)]{Quinlan96} Quinlan, G.\ D.\ 1996, 
New Astron., 1, 35

\bibitem[Rees(1988)]{Rees88}{Rees}, M.\ J.\ 1988, \nat, 333, 523

\bibitem[Richstone et al.(1998)]{Richstone} Richstone, D., et al.\ 1998,
\nat, 395, A14

\bibitem[Sanders(1992)]{Sanders} Sanders, R.\ H.\ 1992, \nat, 359, 131

\bibitem[Sch\"{o}del et al.(2002)]{Schodel02} Sch\"{o}del, R., et al.\ 2002,
\nat, 419, 694

\bibitem[Sesana et al.(2006)]{SHM06} Sesana, A., Haardt,
F., \& Madau, P.\ 2006, \apj, 651, 392

\bibitem[Sridhar \& Tremaine(1992)]{ST92} Sridhar, S., \& Tremaine, 
S.\ 1992, Iarus, 95, 86

\bibitem[Tokovinin et al.(2006)]{Tokovinin06} Tokovinin, A., Thomas, S., Sterzik, M., Udry, S.\ 2006, A\&A, 450, 681

\bibitem[Tremaine et al.(2002)]{Tremaine02} Tremaine, S., et al.\ 2002,
\apj, 574, 740

\bibitem[Volonteri et al.(2003)]{VHM03} Volonteri, M., Haardt, F., \& Madau, P.\ 2003, ApJ, 582, 559

\bibitem[Yu(2002)]{Y02}Yu, Q.\ 2002, \mnras, 331, 935

\bibitem[Yu \& Tremaine(2003)]{YT03}Yu, Q., \& Tremaine, S.\ 2003, \apj, 599, 1129

\end{thebibliography}
\end{document}